\documentclass[aps, prd, reprint, floatfix, superscriptaddress, nofootinbib, showpacs, preprintnumbers, 10pt]{revtex4-1}

\usepackage{amsmath} \usepackage{amssymb} \usepackage{amsfonts}
\usepackage{amsthm} \usepackage{mathtools} \usepackage{mathrsfs}
\usepackage[caption=false, singlelinecheck=false,
justification=raggedright]{subfig} \usepackage{xspace} \usepackage{txfonts}
\usepackage{multirow} \usepackage{array} \usepackage[table]{xcolor}
\usepackage[normalem]{ulem} \usepackage[colorlinks, urlcolor=black,
citecolor=black, link color=black]{hyperref}%
\newcommand{\nubar}{\ensuremath{\bar{\nu}}}
\newcommand{\nul}{\ensuremath{\nu_{\ell}}}
\newcommand{\nulbar}{\ensuremath{\bar{\nu}_{\ell}}}
\newcommand{\Br}{\ensuremath{\text{Br}}}
\newcommand{\few}{\ensuremath{f_{\text{ew}}}}

\newcommand{\nus}{\ensuremath{\nu_{\text{s}}}}

\newcommand{\modulus}[1]{\ensuremath{\left\lvert #1 \right\rvert}}

\newcommand*{\DOI}[2]{\href{http://dx.doi.org/\detokenize{#1}}{#2}}

\allowdisplaybreaks

\begin{document}

\title{Constraints on a sub-eV scale sterile neutrino from non-oscillation measurements} %

\author{C.~S.~Kim} \email[E-mail at: ]{cskim@yonsei.ac.kr}%
\affiliation{Department of Physics and IPAP, Yonsei University, Seoul 120-749,
Korea} %

\author{G. L\'opez Castro}\email[E-mail at: ]{glopez@fis.cinvestav.mx}
\affiliation{Departamento de F\'isica, Centro de Investigaci\'on y de Estudios Avanzados,
Apartado Postal 14-740, 07000 M\'exico Distrito Federal, M\'exico}

\author{Dibyakrupa Sahoo} \email[E-mail at: ]{sahoodibya@yonsei.ac.kr}%
\affiliation{Department of Physics and IPAP, Yonsei University, Seoul 120-749,
Korea} %

\date{\today}

\begin{abstract}
Anomalies in several short-baseline neutrino oscillation experiments suggest the
possible existence of sterile neutrinos at about eV scale having appreciable
mixing with the already known three neutrinos. We find that if such a light
sterile neutrino exists, through a combined study of the leptonic decays of
$\mu^-$, $\tau^-$, $\pi^-$ and $K^-$, some semi-leptonic decays of $\tau^-$ and
the invisible decay width of the $Z$ boson, it is possible to constrain the
relevant mixing matrix elements. Furthermore, we compare the constraints,
derived by using the method presented here, with the experimental results
obtained from short-baseline neutrino oscillation experiments. We find that a
single light sterile neutrino cannot satisfy the existing short-baseline
neutrino oscillation constraints and explain the anomalies mentioned above.
Along the way we provide a number of experimentally clean observables which can
be used to directly study the light sterile neutrino independently of the
neutrino oscillation experiments.
\end{abstract}


\preprint{LDU-2018-01}

\maketitle

\section{Introduction}\label{sec:intro}

Sterile neutrinos, first hypothesized by Pontecorvo~\cite{Pontecorvo:1967fh},
are electrically neutral fermions of either Dirac or Majorana nature with no
standard weak interaction albeit mixing with the existing active neutrinos.
Mathematically, sterile neutrinos are singlets under the gauge symmetry of the
standard model (SM) of particle physics. The theoretical studies of sterile
neutrinos deal with many diverse new physics scenarios which may include a
multitude of sterile neutrinos with masses ranging from below eV scale to close
to the Planck mass scale. In this paper we shall focus only on light sterile
neutrinos which have masses near the eV scale and we discuss how information
from non-oscillation experiments can be used to constrain the mixing matrix
elements of active-sterile neutrino mixing.\footnote{For detailed discussions on
	the kinds of new physics possibilities which include light sterile neutrinos, we
	suggest the reader to look at the reviews in Refs.~\cite{Abazajian:2012ys,
		Gariazzo:2015rra} and the references contained therein.}

The existence of one or more light sterile neutrinos near the eV scale can help
resolve some of the intriguing `anomalies' observed in short-baseline (SBL)
neutrino oscillation experiments, such as the LSND~\cite{LSND-anomaly},
MiniBooNE~\cite{MiniBooNE-anomaly} and Gallium~\cite{Gallium-anomaly}
anomalies.\footnote{It should be noted that the previously known reactor
	neutrino anomaly~\cite{Reactor-anomaly} may not require any explanation in terms
	of light sterile neutrinos in view of the recent paper from the Daya Bay
	collaboration~\cite{An:2017osx}. However, other experiments such as NEOS
	\cite{Ko:2016owz} and DANSS \cite{DANSS} still suggest the presence of this
	reactor neutrino anomaly. For a global analysis of these SBL results we suggest
	the reader to look at Ref.~\cite{SBL:global-analysis}.} In this paper we assume
that there exists only one light sterile neutrino $\nus$ in addition to the
three known active neutrinos ($\nu_e$, $\nu_{\mu}$, $\nu_{\tau}$), all of which
can be written as linear combination of four neutrino mass eigenstates ($\nu_1$,
$\nu_2$, $\nu_3$, $\nu_4$):
\begin{equation}
\nu_{\alpha} = \sum_{i=1}^4 V_{\alpha i} \; \nu _i \ ,
\end{equation}
where $\alpha=e, \mu, \tau, \text{s}$. We assume that the
Pontecorvo-Maki-Nakagawa-Sakata (PMNS)
matrix~\cite{Pontecorvo:1967fh,Maki:1962mu}, the $3\times 3$ matrix that deals
with the mixing of $\nu_e$, $\nu_{\mu}$, $\nu_{\tau}$ with $\nu_1$, $\nu_2$,
$\nu_3$, remains unitary in presence of the sterile neutrino $\nus$, while the
$4\times 4$ mixing matrix $V$ (which might be unrelated to any see-saw mechanism
for generating neutrino mass) can be, in general,
non-unitary~\cite{Cvetic:2017gkt}. In this case, the effects of sterile
neutrinos will become manifest in the observables associated to charged current
interactions of leptons (here repeated labels indicate summation convention,
$\ell = e,\mu,\tau$ and $\widehat{\gamma_{\mu}} \equiv
\gamma_{\mu}(1-\gamma_5)$),
\begin{align}
{\cal L}_W^{\rm CC} &= -\frac{g}{2\sqrt{2}} \, \sum_{\ell = e, \mu, \tau}
\, \left[ \bar{\ell} \; \widehat{\gamma_{\mu}} \, \nu_{\ell} \right] W^{\mu} +
{\rm h.c.} \nonumber\\%
&=  -\frac{g}{2\sqrt{2}} \, \sum_{\ell=e, \mu, \tau} \, \sum_{i=1}^{4} V_{\ell
	i} \left[ \bar{\ell} \; \widehat{\gamma_{\mu}} \, \nu_{i} \right] W^{\mu} + {\rm
	h.c.},\label{eq:ccurrent}
\end{align}
as well as in neutral current $Z$ boson decays into neutrinos,
\begin{align}
{\cal L}_Z^{\rm NC}&= -\frac{g}{4\cos \theta_{W}} \, \sum_{\ell=e, \mu,
	\tau} \, \left[ \bar{\nu}_{\ell} \; \widehat{\gamma_{\mu}} \; \nu_{ \ell}
\right] Z^{\mu} \nonumber \\ %
&=  -\frac{g}{4\cos \theta_{W}} \, \sum_{\ell=e, \mu, \tau} \, \sum_{i=1}^{3}
\bigg[ \bar{\nu}_{i} \; \widehat{\gamma_{\mu}} \, \nu_{i} + \modulus{V_{\ell
		4}}^2 \bar{\nu}_{4} \; \widehat{\gamma_{\mu}} \, \nu_{4}  \nonumber \\%
&\qquad + \left( V_{\ell i}^* V_{\ell 4} \, \bar{\nu}_i \;
\widehat{\gamma_{\mu}} \, \nu_4 + V_{\ell i} V_{\ell 4}^* \, \bar{\nu}_4 \;
\widehat{\gamma_{\mu}} \, \nu_i \right) \bigg] Z^{\mu}, \label{eq:ncurrent}
\end{align}
where $g$ is the weak coupling constant and $\theta_W$ is the weak mixing angle.
The above expression follows from the unitarity of the $3\times 3$ PMNS matrix.
It is important to note that to keep our discussion most general we consider the
$4\times 4$ mixing matrix to be non-unitary. In some new physics scenarios the
sterile neutrino can have a different origin from the active neutrinos, leading
to the non-unitarity of the mixing matrix (for a specific model realizing this
scenario see for instance \cite{Cvetic:2017gkt}). This violation of unitarity,
if observed, would imply presence of unknown new physics.

Taking the Lagrangians of Eqs.~\eqref{eq:ccurrent} and \eqref{eq:ncurrent} into
account we shall explore the effects of this hypothetical light sterile neutrino
in some precision observables and try to set constraints on its mixing with the
known flavor eigenstates. Our purpose is to identify observables that turn out
to be the most sensitive ones and present a clean way to determine the mixing
matrix elements. Given the lightness of this sterile neutrino, its effects on
the different observables considered in this analyses will manifest as an
overall normalization factor. Ratios of decay rates turn out to be very useful
since they are independent of weak couplings, quark mixings and hadronic form
factors. The effects of the sterile neutrino do not cancel in such ratios as
long as they do not satisfy lepton universality, which in our case implies
$\modulus{V_{e4}} \neq \modulus{V_{\mu 4}} \neq \modulus{V_{\tau 4}}$.

Our paper is organized as follows. In Sec.~\ref{sec:weak-decays} we provide a
comprehensive analysis of the relevant weak decays, with particular attention to
the constraints on active-sterile neutrino mixing matrix elements. We do a
combined study of the leptonic decays of $\mu^-$, $\tau^-$, $\pi^-$ and $K^-$,
some semi-leptonic decays of $\tau^-$ as well as the invisible decay of $Z$
boson. We provide all the observables which can be used to constrain the
mentioned mixing matrix elements. In Sec.~\ref{sec:numerical-study} we perform a
numerical study taking all available experimental data and also look for further
predictions which can be tested in oscillation and non-oscillation experiments.
Finally we conclude in Sec.~\ref{sec:conclusion} emphasizing the results and the
uniqueness of our approach.

\section{Probing \texorpdfstring{$\modulus{V_{\ell 4}}$}{Vl4} via weak decays}\label{sec:weak-decays}

Lepton flavor is an absolutely conserved quantum number in the SM with massless
neutrinos. In this limit, we can identify the flavor of neutrinos (or
antineutrinos) produced in processes induced by charged  weak currents by
identifying the flavor of the associated charged lepton, as in the case, for
example, of $\mu^- \to e^- \bar{\nu}_e\nu_{\mu}$ decay. However, if neutrinos
are massive particles, lepton flavor-violating  (LFV) processes like $\mu^- \to
e^- \nul \nulbar$ are possible via a Z-penguin or box diagram at the one loop
level.  Strictly speaking, the observable process when neutrinos are massive is
$\mu^- \to e^- +$ ``missing" because the flavor of neutrinos is not
identified\footnote{In practice, the LFV contribution to the muon rate is
	unobservably small.}. In general, since the light sterile neutrinos, like the
active ones, remain undetected at their place of production, any weak decay of
the type $X \to Y + a \nul + b \nubar_{\ell'}$ is practically $X \to Y +
\text{``missing''}$ where $X$ and $Y$ are some initial and final particle(s)
respectively, $\ell,\ell'=e,\mu,\tau$ and $a,b=0,1$. We shall assume that the
unobserved neutral fermions produced in such weak decays under consideration are
either active or sterile neutrinos (or anti-neutrinos). As we shall show, this
allows us to set bounds on the mixing matrix elements $\modulus{V_{\ell
		4}}$, provided the more sensitive observables to these effects are conveniently
chosen.

\subsection{Leptonic decays of \texorpdfstring{$\mu^-$}{muon} and \texorpdfstring{$\tau^-$} {tau}}

Let us first consider the leptonic $\mu^-$ decay as the reference process. In
the presence of a single sterile neutrino, there are four possible contributions
to muon decay: $\mu^- \to e^-\nubar_e\nu_{\mu},\ e^-\nubar_e\nu_{4}, e^-
\nubar_4\nu_{\mu}$ and $e^- \nubar_4 \nu_4$, all of which contribute to $\mu^-
\to e^- + \text{``missing"}$. The corresponding rate for the $\mu^- \to e^- +
\text{``missing"}$ is given by,
\begin{equation} \label{eq:muon-rate}
\Gamma_{\mu} = \frac{\left(G_F^0\right)^2 }{192\,\pi^3} \; \rho_{\mu e} \;
\Sigma_{\mu e}\ .
\end{equation}
We have defined
\begin{equation}
\rho_{\mu e} = m_{\mu}^5 \, f\left(m_e^2/m_{\mu}^2\right) \, f_W
\left(m_{\mu}\right) \, \few \left(m_{\mu}\right),
\end{equation}
where $m_{\ell}$ denotes the mass of the charged lepton $\ell$, $G_F^0$ is the
Fermi constant if we were to assume no sterile neutrino,
$f(x)=1-8x+8x^3-x^4-12x^2\ln x$, $f_W(m_{\ell})=1+3/5(m_{\ell}/m_W)^2$ is the
finite $W$ mass correction stemming from the $W$ boson propagator, $\few
\left(m_{\ell}\right)$ are the remaining radiative corrections to the decay
rate. Including the effects of the finite mass of electron and the
$O\left(\alpha^2\right)$ radiative corrections, numerically we have: $\few
\left(m_{\mu}\right)=0.995802$~\cite{Tanabashi:2018oca}. The effect of the
sterile neutrino is encoded in the factor $\Sigma_{\mu e}$:
\begin{equation} \label{eq:Sig-m-e}
\Sigma_{\mu e} \equiv \bigg( 1 + \modulus{V_{e4}}^2 \bigg) \left( 1 +
\modulus{V_{\mu 4}}^2 \right).
\end{equation}

The measured value of the effective Fermi constant is given by $G_F = G_F^0
\left(\Sigma_{e\mu}\right)^{1/2}=1.166 3787(6) \times 10^{-5} \text{
	GeV}^{-2}$~\cite{Tanabashi:2018oca}, obtained from a comparison of the muon
decay rate in Eq.~\eqref{eq:muon-rate} and the measured muon lifetime
$\tau_{\mu}=1/\Gamma_{\mu}$. As can be realized, it is not possible to quantify
the effect of sterile neutrino from $\Gamma_{\mu}$ measurement alone without an
independent and precise measurement of $G_F^0$.

A similar expression holds for the decay rates of $\tau^-\to \ell^- +{\rm
	``missing"}$ decays (with $\ell=e, \mu$), which in the presence of an additional
sterile neutrino becomes:
\begin{equation} \label{eq:tau-rate}
\Gamma_{\tau}^{\ell} = \frac{\left(G_F^0\right)^2}{192 \, \pi^3} \; \rho_{\tau
	\ell} \; \Sigma_{\tau\ell}\ ,
\end{equation}
with
\begin{equation}
\rho_{\tau \ell} = m_{\tau}^5 \, f\left(m_{\ell}^2/m_{\tau}^2\right) \, f_W
\left(m_{\tau}\right) \, \few^{\ell} \left(m_{\tau}\right),
\end{equation}
where we have a similar expression for the $\Sigma_{\tau\ell}$ factor as in
Eq.~\eqref{eq:Sig-m-e}, under the corresponding replacement of flavor indices.
The numerical values of the radiative corrections are
$\few^e\left(m_{\tau}\right) = 0.995722$ and $\few^{\mu}\left(m_{\tau}\right) =
0.995960$~\cite{Tanabashi:2018oca}. If we compute the ratio between
Eqs.~\eqref{eq:muon-rate} and \eqref{eq:tau-rate}, we obtain
\begin{equation} \label{eq:R-tau-mu}
\frac{\Gamma_{\tau}^{\ell}}{\Gamma_{\mu}} = \frac{\rho_{\tau \ell}}{\rho_{\mu
		e}} \; \frac{\Sigma_{\tau \ell}}{\Sigma_{\mu e}}\ .
\end{equation}
If we take the ratio of $\Gamma_{\tau}^{e}$ and $\Gamma_{\tau}^{\mu}$ using
Eq.~\eqref{eq:tau-rate}, we get,
\begin{equation}\label{eq:tau-mu-e}
\frac{\Gamma_{\tau}^e}{\Gamma_{\tau}^{\mu}} = \left(
\frac{\rho_{\tau e}}{\rho_{\tau \mu}}
\right) \left( \frac{1 + \modulus{V_{e 4}}^2}{1 + \modulus{V_{\mu 4}}^2}
\right).
\end{equation}
It is clear from Eqs.~\eqref{eq:R-tau-mu} and \eqref{eq:tau-mu-e} that if there
is no lepton universality, i.e.\ $\modulus{V_{e4}} \neq \modulus{V_{\mu 4}} \neq
\modulus{V_{\tau 4}}$, we can find out some observables which can probe the
active-sterile mixing, without being worried about the extraction of the Fermi
constant $G_F^0$. Since all the mixing matrix elements $\modulus{V_{\ell
		4}}^2$ are positive and do not exceed unity, we have $1/2 \leq
\Sigma_{\tau\ell}/\Sigma_{\mu e}, \Sigma_{\tau e}/\Sigma_{\tau\mu} \leq 2$. In
practice, given the good agreement of the SM with experimental data for the
leptonic decays of $\tau$, we would expect to have
$\Sigma_{\tau\ell}/\Sigma_{\mu e}$ very close to $1$.

Note that we can always express the partial decay rates of the decays of
$\tau^-$ in terms of branching ratios (denoted by $\Br\left(\tau^- \to \ell^- +
\text{``missing''}\right)$) and mean lifetime of $\tau^-$ (denoted by
$\tau_{\tau}$): $\Gamma_{\tau}^{\ell} = \Br\left(\tau^- \to \ell^- +
\text{``missing''}\right)/\tau_{\tau}$. In this way we obtain the following
observables, which are ratios involving the mixing matrix elements
$\modulus{V_{\ell 4}}$,
\begin{subequations}\label{eq:ratio-observables}
\begin{align}
R_{\tau/e} &\equiv \frac{1 + \modulus{V_{\tau 4}}^2}{1 + \modulus{V_{e4}}^2} =
\frac{\Br\left( \tau \to \mu + \text{``missing''} \right)}{\Br\left( \mu \to e +
	\text{``missing''} \right)} \frac{\tau_{\mu}}{\tau_{\tau}} \frac{\rho_{\mu
		e}}{\rho_{\tau \mu}}\ ,\label{eq:R-t-e}\\%
R_{\tau/\mu} &\equiv \frac{1 + \modulus{V_{\tau 4}}^2}{1 + \modulus{V_{\mu
			4}}^2} = \frac{\Br\left( \tau \to e + \text{``missing''} \right)}{\Br\left( \mu
	\to e + \text{``missing''} \right)} \frac{\tau_{\mu}}{\tau_{\tau}}
\frac{\rho_{\mu e}}{\rho_{\tau e}}\ ,\label{eq:R-t-m}\\%
R_{e/\mu} &\equiv \frac{1 + \modulus{V_{e 4}}^2}{1 + \modulus{V_{\mu 4}}^2} =
\frac{\Br\left( \tau \to e + \text{``missing''}\right)}{\Br\left( \tau \to \mu +
	\text{``missing''}\right)} \frac{\rho_{\tau \mu}}{\rho_{\tau e}}\
.\label{eq:R-e-m}
\end{align}
\end{subequations}
These three observables are not independent, by definition, since $R_{\tau/e} \,
R_{e/\mu} = R_{\tau/\mu}$. Therefore, only two out of these three ratios would
be useful for our numerical analysis and we would need some extra independent
observable(s) in order to constrain the three active-sterile mixing
matrix elements.

It is important to note that the ratio observables $R_{\tau/e}$, $R_{e/\mu}$ and
$R_{\tau/\mu}$ actually probe the unitarity of the $4\times4$ mixing matrix. If
we were to relax the assumption that the $3\times3$ PMNS matrix is unitary, the
ratio observables take the form
\begin{equation}\label{eq:gen-ratio-observable}
R_{\ell/\ell'} = \left(\sum_{i=1}^4 \modulus{V_{\ell i}}^2\right)/\left(\sum_{j=1}^4
	\modulus{V_{\ell' j}}^2\right),
\end{equation}
where $\ell \neq \ell'$ and $\ell, \ell'=e,\mu,\tau$. It is clear from
Eq.~\eqref{eq:gen-ratio-observable} that $R_{\ell/\ell'} = 1$ only when both
$\sum_{i=1}^4 \modulus{V_{\ell i}}^2=1$ and $\sum_{j=1}^4
\modulus{V_{\ell'j}}^2~=~1$. It should be noted that if one were to consider an
$n \times n$ active-sterile mixing matrix with $n>4$ in some new physics model,
then the summations in Eq.~\eqref{eq:gen-ratio-observable} would run from $1$ to
$n$. Thus, $R_{\ell/\ell'}$ can also probe the unitarity of such an
active-sterile mixing matrix in the most general scenario.\footnote{It is
	important to note that here we are not considering the possibility of any
	sterile neutrinos being kinematically inaccessible to our decay modes under
	consideration. If such heavy sterile neutrinos are experimentally found to
	exist, it is beyond the scope of our analysis and $R_{\ell/\ell'}$ can not probe
	unitarity of the full active-sterile neutrino mixing matrix.}

\subsection{Semileptonic decays of \texorpdfstring{$\tau^-$}{tau} and leptonic decays of \texorpdfstring{$\pi^-$}{pi} and \texorpdfstring{$K^-$}{K}}

Ratios of semileptonic decays of $\tau^-$ and leptonic decays of $\pi^-$ and
$K^-$ can also be useful to constrain the mixings of a light sterile neutrino.
Let us consider the $\tau^-\to P^-\nu_{\tau}$ and $P^- \to \ell^-\nubar_{\ell}$
decays ($P=\pi$ or $K$ mesons and $\ell=e$ or $\mu$), which are the most
precisely measured processes of this type. In the presence of a light sterile
neutrino, the decay rates of these processes are given by:
\begin{equation}\label{eq:t2Pn}
\Gamma_{\tau \to P \nu} = \frac{\left(G_F^0\right)^2 \modulus{V_{uq}}^2}{16\pi}
f_P^2 \, m_{\tau}^3 \left( 1 - \frac{m_P^2}{m_{\tau}^2} \right)^2
\delta_{\tau}^P \left( 1 + \modulus{V_{\tau 4}}^2 \right)\ ,
\end{equation}
and
\begin{equation}\label{eq:P2ln}
\Gamma_{P \to \ell \nu} = \frac{\left(G_F^0\right)^2 \modulus{V_{uq}}^2}{8\pi}
f_P^2 \, m_{\ell}^2 \, m_P \left(1-\frac{m_{\ell}^2}{m_{P}^2} \right)^2
\delta_{P}^{\ell} \left( 1 + \modulus{V_{\ell 4}}^2 \right)\  .
\end{equation}
In the above expressions, $f_P$ denotes the decay constant of the $P^-$ meson,
which can  be calculated from Lattice QCD,  $V_{uq}$ (with $q=d(s)$ for $P=\pi
(K)$ mesons) is the Cabibbo-Kobayashi-Maskawa (CKM) matrix element and
$\delta_{\tau}^P$ (or $\delta_{P}^{\ell}$) denotes the radiative corrections to
$\tau\to P \nu$ (or $P\to \ell \nu$) decay.

The ratio of the these decays,
\begin{equation} \label{eq:tau-pi-k}
\frac{\Gamma_{\tau \to P \nu }}{\Gamma_{P \to \ell \nu}} =
\frac{m_P^3}{2m_{\ell}^2m_{\tau}} \left(
\frac{m_{\tau}^2-m_P^2}{m_P^2-m_{\ell}^2}\right)^2 \left(
\frac{\delta_{\tau}^P}{\delta_{P}^{\ell}}\right) \left( \frac{1 +
	\modulus{V_{\tau 4}}^2}{1 + \modulus{V_{\ell 4}}^2} \right)\ ,
\end{equation}
is independent of the parameters related to the hadronic vertex and can provide
a clean information about the active-sterile mixing matrix elements
provided there is no lepton universality. Furthermore, the ratio of electron and
muon channels in $P$ meson decays,
\begin{equation} \label{eq:P-e-mu}
\frac{\Gamma_{P \to e\nu}}{\Gamma_{P \to \mu \nu}} = \frac{m_e^2}{m_{\mu}^2}
\left(\frac{m_P^2-m_e^2}{m_P^2-m_{\mu}^2}\right)^2
\left(\frac{\delta_P^e}{\delta_P^{\mu}}\right) \left( \frac{1 +
	\modulus{V_{e4}}^2}{1 + \modulus{V_{\mu 4}}^2} \right)\ ,
\end{equation}
can also be used to study the active-sterile mixing matrix elements since
it is independent of hadronic inputs and common terms of radiative corrections
also cancel in the ratio.

The radiative corrections can be split into a short-distance (SD) and a
long-distance (LD) parts: $\delta=\delta(SD)+\delta(LD)$. The dominant
contribution to SD corrections in $\tau$ and $P$ meson decays is given by
\begin{equation}
\delta_{X}^Y(SD)=1+\frac{2\alpha}{\pi} \ln \left(\frac{m_Z}{m_X} \right)\ .
\end{equation}
The computation of the LD parts are more difficult to evaluate, since they
depend on the details of the strong interactions in the transition regime (1-2
GeV), which involve usage of phenomenological models including contributions
from possible resonances as well as scalar QED, so that one can consider the
meson-photon interactions beyond the point-meson approximation
\cite{LD:difficulty, Decker:1994ea, Cirigliano:2007ga}. For numerical analysis
we shall use the ratio of radiative corrections in the semileptonic $\tau$
decays as given in Ref.~\cite{Decker:1994ea}:
\begin{equation}
\frac{\delta_{\tau}^P}{\delta_{P}^{\mu}} =
\begin{cases}
1.0016 \pm 0.0014 , & \text{ for } P=\pi^-\\%
1.0090 \pm 0.0022 , & \text{ for } P=K^-
\end{cases} \ .
\end{equation}
Similarly, the ratio of $O(\alpha)$ radiative corrections for $P$ meson decays
is given by \cite{Cirigliano:2007ga}:
\begin{equation}
\frac{\delta_P^e}{\delta_P^{\mu}} =
\begin{cases}
0.9625 \pm 0.0001, & \text{ for } P=\pi^- \\%
0.9642 \pm 0.0004, & \text{ for } P=K^-
\end{cases}\ .
\end{equation}
Once again, replacing the partial decay rates in terms of branching ratios and
lifetimes we can define, by using Eqs.~\eqref{eq:tau-pi-k} and
\eqref{eq:P-e-mu}, the equivalent ratio observables as in
Eq.~\eqref{eq:ratio-observables}:
\begin{subequations}\label{eq:ratio-observables-2}
\begin{align}
R_{\tau/e} &= \frac{\Br\left( \tau \to P\nu \right)}{\Br\left( P \to e \nu
	\right)} \frac{\tau_{P}}{\tau_{\tau}} \frac{2 m_{e}^2 m_{\tau}}{m_P^3} \left(
\frac{m_P^2-m_{e}^2}{m_{\tau}^2-m_P^2}\right)^2 \left(
\frac{\delta_{P}^{e}}{\delta_{\tau}^P}\right)\ ,\\%
R_{\tau/\mu} &= \frac{\Br\left( \tau \to P\nu \right)}{\Br\left( P \to \mu \nu
	\right)} \frac{\tau_{P}}{\tau_{\tau}} \frac{2 m_{\mu}^2 m_{\tau}}{m_P^3}
\left(\frac{m_P^2-m_{\mu}^2}{m_{\tau}^2-m_P^2}\right)^2 \left(
\frac{\delta_{P}^{\mu}}{\delta_{\tau}^P}\right)\ ,\label{eq:Rtm}\\%
R_{e/\mu} &= \frac{\Br\left( P \to e \nu\right)}{\Br\left( P \to \mu \nu\right)}
\frac{m_{\mu}^2}{m_e^2} \left(\frac{m_P^2-m_{\mu}^2}{m_P^2-m_e^2}\right)^2
\left(\frac{\delta_P^{\mu}}{\delta_P^e}\right) \ .\label{eq:Rem}
\end{align}
\end{subequations}
It must be noted that each ratio defined above has got two values, one
corresponding to $P=\pi$ and the other for $P=K$. Even though we did not get any
extra observables here, we can probe the same observables from a different set
of decays than the purely leptonic decays of $\mu$ and $\tau$ as given in
Eq.~\eqref{eq:ratio-observables}. In fact, if we consider only those decays that
are mediated by charged current interaction, we shall be constrained to consider
decays of the type $X \to Y + a \nu_{\ell} + b \nubar_{\ell'}$, with
$\ell,\ell'=e,\mu,\tau$ and $\ell \neq \ell'$, $a,b=0,1$ and $X$, $Y$ being
appropriate particles. From such decays we can extract only the ratios
$R_{\tau/e}$, $R_{\tau/\mu}$ and $R_{e/\mu}$ and nothing more unless we use some
other independent information such as $G_F^0$ or hadronic form factors or CKM
matrix elements. For this reason, considering the weak neutral current processes
can be useful.

\subsection{Invisible width of the $Z$ boson and the number of light active neutrinos}

In the presence of a light sterile neutrino, the contributions to the
`invisible' decay width of the $Z$ gauge boson that stem from
Eq.~\eqref{eq:ncurrent} are: $\nulbar \nul, \nulbar \nu_4, \nubar_4 \nul$ and
$\nubar_4 \nu_4$. The invisible width of $Z$ is, therefore, given by:
\begin{equation}\label{eq:Z-invisible-width}
\Gamma_{\text{inv}} = \frac{G_F^0 m_Z^3}{12\sqrt{2}\,\pi} \sum_{\ell=e,\mu,\tau}
\left( 1 + \modulus{V_{\ell 4}}^2 \right)^2,
\end{equation}
where $m_Z$ denotes the mass of the $Z$ boson. It is interesting to note that
the Fermi constant used in Eq.~\eqref{eq:Z-invisible-width} is extracted from
the muon decay, which under our assumption of one sterile neutrino leads to
Eq.~\eqref{eq:muon-rate}. Therefore, in terms of the measured Fermi constant
$G_F$, the invisible width of $Z$ is given by,
\begin{equation}\label{eq:Z-invisible-width-2}
\Gamma_{\text{inv}} = \frac{G_F\,m_Z^3}{12 \sqrt{2} \,\pi}
\frac{1}{\sqrt{\Sigma_{\mu e}}} \sum_{\ell=e,\mu,\tau} \left( 1 +
\modulus{V_{\ell 4}}^2 \right)^2.
\end{equation}
If we consider $V_{\ell 4} =0$, we get the expression for
$\Gamma_{\text{inv}}^{\text{SM}}$ which is the invisible width of the $Z$ boson
in the SM. It is well known that the number of light neutrinos ($N_{\nu}$) is
extracted from observed invisible width of $Z$ boson by using the expression,
\begin{equation}
\frac{N_{\nu}}{3} = \frac{\Gamma_{\text{inv}}}{\Gamma_{\text{inv}}^{\text{SM}}}.
\end{equation}
Thus, we can express $N_{\nu}$ in terms of the active-sterile mixing matrix
elements as follows,
\begin{equation}\label{eq:Nnu}
N_{\nu} = \frac{1}{\sqrt{\Sigma_{\mu e}}} \sum_{\ell=e,\mu,\tau} \left( 1 +
\modulus{V_{\ell 4}}^2 \right)^2.
\end{equation}
It is clear that $N_{\nu}$ can now be used in conjunction with the ratio
operators $R_{\tau/e}$, $R_{\tau/\mu}$ and $R_{e/\mu}$ to constrain the
active-sterile mixing matrix elements. From Refs.~\cite{Tanabashi:2018oca,
	ALEPH:2005ab} we get the number of light neutrinos to be $N_{\nu}=2.9840 \pm
0.0082$.

\subsection{Energy spectrum of charged lepton in leptonic tau decay}

\begin{figure}[hbtp]
\centering %
\includegraphics[width=0.9\linewidth,keepaspectratio]{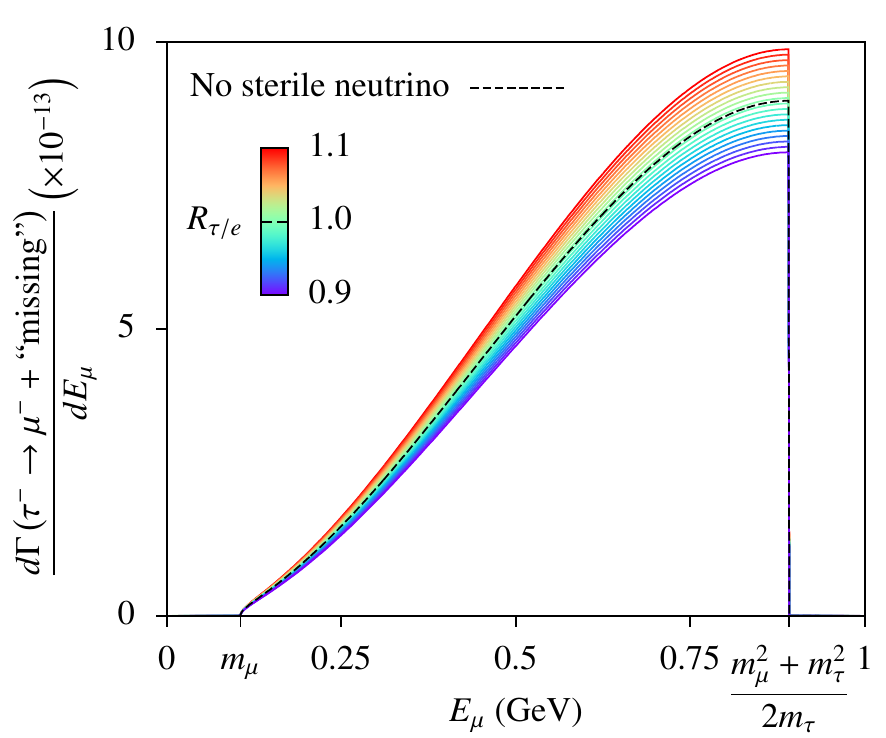} %
\caption{Muon energy spectrum in $\tau^- \to \mu^- + \text{``missing''}$ decays
	for values of $R_{\tau/e} = \Sigma_{\tau\mu}/\Sigma_{\mu e}$ in the range
	$(0.9,1.1)$ using Eq.~\eqref{eq:E-spectrum}. We have considered lepton
	non-universality here, i.e. $\modulus{V_{e4}} \neq \modulus{V_{\mu 4}} \neq
	\modulus{V_{\tau 4}}$, as otherwise $R_{\tau/e} =1$ even for the case of one
	sterile neutrino.}%
\label{fig:sample-spectrum}
\end{figure}

Till now we have considered fully phase-space integrated partial decay rates for
some well chosen decays as a means to constrain the active-sterile mixing.
It is interesting to ask whether differential partial decay rates or
distributions can be used to look for signatures of the sterile neutrino. In
presence of one light sterile neutrino, the energy distribution of the final
charged lepton $\ell$ in the decay $\tau \to \ell + \text{``missing''}$ gets
modified as follows,
\begin{align}
\frac{d\Gamma\left(\tau \to \ell + \text{``missing''}\right)}{dE_{\ell}}
&=\left(\frac{32m_{\tau}}{\rho'_{\mu e} \,\tau_{\mu}}\right)
\left(3E_{\ell}E_{\ell}^{\rm max}-m_{\ell}^2-2E_{\ell}^2 \right) \nonumber \\%
& \qquad \times \sqrt{E_{\ell}^2-m_{\ell}^2} \left(
\frac{\Sigma_{\tau\ell}}{\Sigma_{\mu e}} \right)\ ,\label{eq:E-spectrum}
\end{align}
where $\rho'_{\mu e} = m_{\mu}^5 \, f\left(m_e^2/m_{\mu}^2\right)$ and
$E_{\ell}^{\rm max}=(m_{\tau}^2+m_{\ell}^2)/2m_{\tau}$ with $\ell=e,\mu$. This
expression does not contain the effects  of radiative corrections to the lepton
spectrum. It is important to note that this is not a normalized distribution
since in that case the effect of sterile neutrinos will cancel during
normalization. In Fig.~\ref{fig:sample-spectrum} we plot the energy distribution
of the muon in $\tau^- \to \mu^- + {\rm missing}$ decays. We have chosen $0.9
\leq \Sigma_{\tau\mu}/\Sigma_{\mu e} \leq 1.1$ as an example for the range of
the free parameter. The effect of the sterile neutrino is the same over the
whole spectrum although it is more discernible at the end-point of the energy
spectrum. %
It is important to note that in order to get a prediction for the effect of
sterile neutrino on the energy spectrum as per the present data, we must first
constrain the individual mixing matrix elements using our set of
observables.

\subsection{Analytical solutions for \texorpdfstring{$\modulus{V_{\ell 4}}^2$}{|Vl4|**2} in terms of observables}

We have four observables ($R_{\tau /e}$, $R_{\tau/\mu}$, $R_{e/\mu}$ and
$N_{\nu}$) which can be defined in terms of three active-sterile mixing
matrix elements ($\modulus{V_{e4}}^2$, $\modulus{V_{\mu 4}}^2$ and
$\modulus{V_{\tau 4}}^2$) as shown in Eqs.~\eqref{eq:ratio-observables} and
\eqref{eq:Nnu}. Since the three ratio observables of
Eq.~\eqref{eq:ratio-observables} are not independent, we can always consider
$N_{\nu}$ and any two out of the three ratios to solve for the mixing
matrix elements. This gives rise to three closely related schemes of
analytical solutions. In Sec.~\ref{sec:numerical-study} we shall present our
numerical analysis following all the three schemes.

\textbf{Scheme A:} In this scheme we shall use $R_{\tau/\mu}$, $R_{e/\mu}$ and
$N_{\nu}$ to solve for the mixing matrix elements. From
Eqs.~\eqref{eq:Sig-m-e}, \eqref{eq:R-t-m} and \eqref{eq:R-e-m} it is
straightforward to get,
\begin{subequations}\label{eq:Nnu-sub}
	\begin{align}
	\Sigma_{\mu e} &= R_{e/\mu} \left(1 + \modulus{V_{\mu 4}}^2\right)^2,\\%
	\modulus{V_{\tau 4}}^2 &= R_{\tau/\mu} \left(1 + \modulus{V_{\mu 4}}^2\right)
	-1, \label{eq:V-t4-sq}\\%
	\modulus{V_{e 4}}^2 &= R_{e/\mu} \left(1 + \modulus{V_{\mu 4}}^2\right) -1.
	\label{eq:V-e4-sq}
	\end{align}
\end{subequations}
Substituting Eq.~\eqref{eq:Nnu-sub} in Eq.~\eqref{eq:Nnu} and simplifying we get
\begin{equation}\label{eq:V-m4-sq-1}
\modulus{V_{\mu 4}}^2 = \frac{N_{\nu} \sqrt{R_{e/\mu}}}{1 + R_{e/\mu}^2 +
	R_{\tau/\mu}^2} - 1.
\end{equation}
Finally substituting Eq.~\eqref{eq:V-m4-sq-1} in Eqs.~\eqref{eq:V-t4-sq} and
\eqref{eq:V-e4-sq} we get,
\begin{subequations}\label{eq:V-t4Sq-e4Sq}
	\begin{align}
	\modulus{V_{\tau 4}}^2 = \frac{N_{\nu} \, R_{\tau/\mu} \sqrt{R_{e/\mu}}}{1 +
		R_{e/\mu}^2 + R_{\tau/\mu}^2} - 1,\label{eq:V-t4-sq-1}\\%
	\modulus{V_{e 4}}^2 = \frac{N_{\nu} \, R_{e/\mu}^{3/2}}{1 + R_{e/\mu}^2 +
		R_{\tau/\mu}^2} - 1.\label{eq:V-e4-sq-1}
	\end{align}
\end{subequations}
Thus Eqs.~\eqref{eq:V-m4-sq-1} and \eqref{eq:V-t4Sq-e4Sq} express the
active-sterile mixing matrix elements in terms of observables
$R_{\tau/\mu}$, $R_{e/\mu}$ and $N_{\nu}$.

\textbf{Scheme B:} In this scheme we shall use the observables $R_{\tau/e}$,
$R_{e/\mu}$ and $N_{\nu}$. The expressions for active-sterile mixing matrix
elements in this scheme are obtained from Eqs.~\eqref{eq:V-m4-sq-1} and
\eqref{eq:V-t4Sq-e4Sq} by making use of the identity $R_{\tau/\mu} = R_{\tau/e}
\, R_{e/\mu}$. Thus, in this scheme we have,
\begin{subequations}
	\begin{align}
	\modulus{V_{e 4}}^2 = \frac{N_{\nu} \, R_{e/\mu}^{3/2}}{1 + R_{e/\mu}^2 \left( 1
		+ R_{\tau/e}^2 \right)} - 1,\label{eq:V-e4-sq-2}\\%
	\modulus{V_{\mu 4}}^2 = \frac{N_{\nu} \sqrt{R_{e/\mu}}}{1 + R_{e/\mu}^2 \left( 1
		+ R_{\tau/e}^2 \right)} - 1,\label{eq:V-m4-sq-2}\\%
	\modulus{V_{\tau 4}}^2 = \frac{N_{\nu} \, R_{\tau/e} \, R_{e/\mu}^{3/2}}{1 +
		R_{e/\mu}^2 \left( 1 + R_{\tau/e}^2 \right)} - 1.\label{eq:V-t4-sq-2}
	\end{align}
\end{subequations}

\textbf{Scheme C:} In this scheme we shall use the observables $R_{\tau/e}$,
$R_{\tau/\mu}$ and $N_{\nu}$. The expressions for active-sterile mixing
matrix elements in this scheme are obtained from Eqs.~\eqref{eq:V-m4-sq-1}
and \eqref{eq:V-t4Sq-e4Sq} by making use of the identity $R_{e/\mu} =
R_{\tau/\mu} / R_{\tau/e}$. Thus, in this scheme we have,
\begin{subequations}
	\begin{align}
	\modulus{V_{e 4}}^2 = \frac{N_{\nu} \, R_{\tau/\mu}^{3/2}
		\sqrt{R_{\tau/e}}}{R_{\tau/e}^2 + R_{\tau/\mu}^2 + R_{\tau/e}^2 \,
		R_{\tau/\mu}^2} - 1,\label{eq:V-e4-sq-3}\\%
	\modulus{V_{\mu 4}}^2 = \frac{N_{\nu} \, R_{\tau/e}^{3/2}
		\sqrt{R_{\tau/\mu}}}{R_{\tau/e}^2 + R_{\tau/\mu}^2 + R_{\tau/e}^2 \,
		R_{\tau/\mu}^2} - 1,\label{eq:V-m4-sq-3}\\%
	\modulus{V_{\tau 4}}^2 = \frac{N_{\nu} \, R_{\tau/e}^{3/2}
		R_{\tau/\mu}^{3/2}}{R_{\tau/e}^2 + R_{\tau/\mu}^2 + R_{\tau/e}^2 \,
		R_{\tau/\mu}^2} - 1.\label{eq:V-t4-sq-3}
	\end{align}
\end{subequations}

It is important to note that the ratio observables can be determined by using
either Eq.~\eqref{eq:ratio-observables} or \eqref{eq:ratio-observables-2}. In
the numerical analysis ahead in Sec.~\ref{sec:numerical-study} we shall consider
both these options. Once we know the values of $\modulus{V_{\ell 4}}^2$ it would
be interesting to know what would be their impact on short-baseline neutrino
oscillation experiments.

\subsection{Impact of \texorpdfstring{$\modulus{V_{\ell 4}}$}{|Vl4|} on short-baseline neutrino oscillation}

In short-baseline (SBL) neutrino oscillation experiments, where the
active-sterile neutrino oscillations are easier to observe, the effective
probability of neutrino oscillations from an initial flavor state $\nu_{\alpha}$
to a final flavor state $\nu_{\beta}$ is given by
\begin{equation}
P_{\alpha\beta}^{\text{(SBL)}} \simeq \modulus{\delta_{\alpha\beta} - \sin^2
	2\theta_{\alpha\beta} \, \sin^2 \left(\frac{\Delta m_{\text{SBL}}^2 \,
		L}{4E}\right)},
\end{equation}
where $L$ is the distance between the neutrino source and the detector, $E$ is
the energy of the neutrino beam, $\Delta m_{\text{SBL}}^2$ is the new
squared-mass difference corresponding to oscillations between the sterile and
active neutrinos, and the oscillation amplitude is given by,
\begin{equation}
\sin^2 2\theta_{\alpha\beta} = 4 \modulus{V_{\alpha 4}}^2
\modulus{\delta_{\alpha\beta} - \modulus{V_{\beta 4}}^2}.
\end{equation}
We are interested in $\sin^2 2\theta_{\mu e}$, $\sin^2 2\theta_{ee}$ and $\sin^2
2\theta_{\mu\mu}$ which can be easily constrained once we know
$\modulus{V_{e4}}^2$ and $\modulus{V_{\mu 4}}^2$ by using our schemes A, B, C as
discussed before.

\section{Numerical analysis and discussion}\label{sec:numerical-study}

\begin{figure}[hbtp]
	\centering%
\includegraphics[width=\linewidth,keepaspectratio]{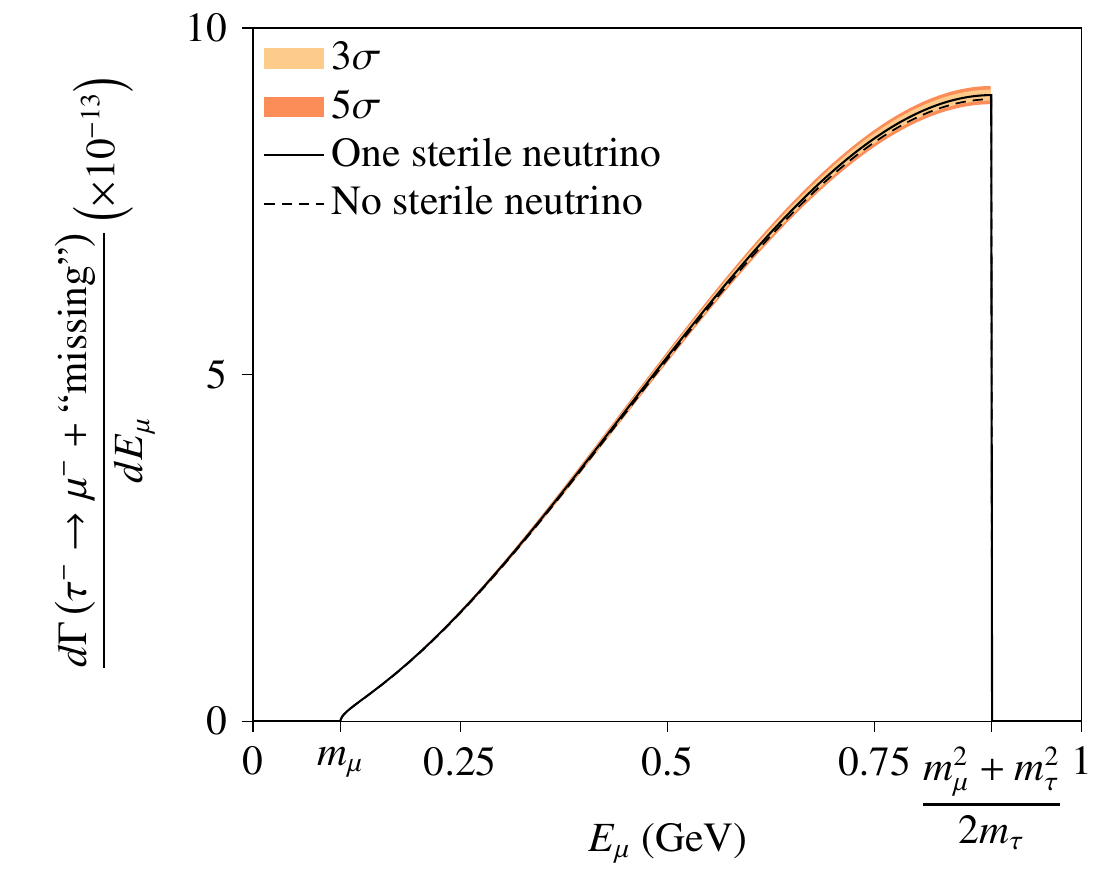}%
\caption{The unnormalized muon energy distribution in the decay $\tau^- \to
	\mu^- + \text{``missing''}$ considering $\Sigma_{\tau\mu}/\Sigma_{\mu e} =
	R_{\tau/e}$ from Eq.~\eqref{eq:Rte-val}.}%
\label{fig:distribution}
\end{figure}

In order to numerically estimate the active-sterile mixing matrix elements using
the expressions obtained in the previous section under schemes A, B and C, we
have used all the precise experimental results for branching ratios, lifetimes,
masses, radiative corrections and the number of light neutrinos $N_{\nu}$ as
reported by the Particle Data Book~\cite{Tanabashi:2018oca}. In the numerical
study we have used Eq.~\eqref{eq:ratio-observables}, \eqref{eq:Rtm} and
\eqref{eq:Rem} for the ratio observables considering both $P=\pi$ and $P=K$. We
have done a simple propagation of errors following the method of quadrature. Our
numerical calculation gives the following values for the ratio observables,
\begin{subequations}\label{eq:ratio-values}
\begin{align}
R_{\tau/e} &= \ \ 1.00667 \pm 0.00293 \quad \text{(Using
	Eq.~\eqref{eq:R-t-e})},\label{eq:Rte-val}\\%
R_{\tau/\mu} &=
\begin{cases}
1.00290 \pm 0.00287 & \text{(Using Eq.~\eqref{eq:R-t-m})}\\%
0.99406 \pm 0.00594 & \text{(Using Eq.~\eqref{eq:Rtm} for $P=\pi$)}\\%
0.97807 \pm 0.01443 & \text{(Using Eq.~\eqref{eq:Rtm} for $P=K$)}
\end{cases}, \label{eq:Rtm-val}\\%
R_{e/\mu} &=
\begin{cases}
0.99626 \pm 0.00320 & \text{(Using Eq.~\eqref{eq:R-e-m})}\\%
0.99589 \pm 0.00324 & \text{(Using Eq.~\eqref{eq:Rem} for $P=\pi$)}\\%
1.00436 \pm 0.00479 & \text{(Using Eq.~\eqref{eq:Rem} for $P=K$)}
\end{cases}. \label{eq:Rem-val}
\end{align}
\end{subequations}
It is quite clear from these results that all the estimates for the ratio
observables are consistent with $1$ within $2\sigma$ standard deviations. This
implies that the $4\times4$ mixing matrix is unitary irrespective of whether we
consider the $3\times3$ PMNS matrix to be unitary or not. Using the value of
$R_{\tau/e}$ from Eq.~\eqref{eq:Rte-val} in Eq.~\eqref{eq:E-spectrum} we can
plot the unnormalized energy distribution of the muon in the decay $\tau^- \to
\mu^- + \text{``missing''}$, as shown in Fig.~\ref{fig:distribution}. It is
clear from this distribution that the current data from weak decays is
consistent with absence of light sterile neutrinos. Since the search for light
sterile neutrinos is traditionally done via short-baseline neutrino oscillation
experiments, looking at the estimates of active-sterile mixing matrix elements
as well as the oscillation amplitude would be very useful.

\begin{table*}[hbtp]
\centering%
\subfloat[]{\label{tab:Ve4Sq}%
\fbox{%
\begin{tabular}{|c|c|c|c|c|c|c|}
\multicolumn{7}{c}{\textbf{Predicted values for $\modulus{V_{e4}}^2$}}\\ \hline%
\multicolumn{3}{|c}{\cellcolor{gray!20}\textbf{Scheme A}} & \multicolumn{3}{c|}{ } & \cellcolor{gray!20}\textbf{Scheme C} \\ \cline{4-7}%
\multicolumn{3}{|c|}{ } & \multicolumn{3}{c|}{$R_{e/\mu}$} & $R_{\tau/e}$\\ \cline{4-7}%
\multicolumn{3}{|c|}{ } & \multirow{2}{*}{Eq.~\eqref{eq:R-e-m}} & \multicolumn{2}{c|}{Eq.~\eqref{eq:Rem}} & \multirow{2}{*}{Eq.~\eqref{eq:R-t-e}} \\ \cline{5-6}%
\multicolumn{3}{|c|}{ } & & $P=\pi$ & $P=K$ & \\ \hline%
\multirow{3}{*}{$R_{\tau/\mu}$} & \multicolumn{2}{c|}{Eq.~\eqref{eq:R-t-m}} & $0.01036 \pm 0.00425$ & $0.01067 \pm 0.00428$ & $0.00365 \pm 0.00517$ & $0.01036 \pm 0.00276$ \\ \cline{2-7}%
& \multirow{2}{*}{Eq.~\eqref{eq:Rtm}} & $P=\pi$ & $0.00450 \pm 0.00549$ & $0.00481 \pm 0.00551$ & $0.00222 \pm 0.00623$ & $0.01186 \pm 0.00290$ \\ \cline{3-7}%
& & $P=K$ & $0.00614 \pm 0.01037$ & $0.00583 \pm 0.01038$ & $0.01288 \pm 0.01079$ & $0.01478 \pm 0.00386$ \\ \hline%
\multicolumn{3}{c}{\cellcolor{gray!20}\textbf{Scheme B}} & \multicolumn{4}{c}{ } \\ \cline{1-6}%
\multicolumn{3}{|c|}{$R_{\tau/e}$} & $0.01036 \pm 0.00277$ & $0.01043 \pm 0.00277$ & $0.00905 \pm 0.00283$ & \multicolumn{1}{c}{ }\\ \cline{1-6}%
\end{tabular}%
}}\\%
\subfloat[]{\label{tab:Vm4Sq}%
\fbox{%
\begin{tabular}{|c|c|c|c|c|c|c|}
\multicolumn{7}{c}{\textbf{Predicted values for $\modulus{V_{\mu 4}}^2$}}\\ \hline%
\multicolumn{3}{|c}{\cellcolor{gray!20}\textbf{Scheme A}} & \multicolumn{3}{c|}{ } & \cellcolor{gray!20}\textbf{Scheme C} \\ \cline{4-7}%
\multicolumn{3}{|c|}{ } & \multicolumn{3}{c|}{$R_{e/\mu}$} & $R_{\tau/e}$\\ \cline{4-7}%
\multicolumn{3}{|c|}{ } & \multirow{2}{*}{Eq.~\eqref{eq:R-e-m}} & \multicolumn{2}{c|}{Eq.~\eqref{eq:Rem}} & \multirow{2}{*}{Eq.~\eqref{eq:R-t-e}} \\ \cline{5-6}%
\multicolumn{3}{|c|}{ } & & $P=\pi$ & $P=K$ & \\ \hline%
\multirow{3}{*}{$R_{\tau/\mu}$} & \multicolumn{2}{c|}{Eq.~\eqref{eq:R-t-m}} & $0.00665 \pm 0.00337$ & $0.00659 \pm 0.00337$ & $0.00798 \pm 0.00342$ & $0.00665 \pm 0.00361$ \\ \cline{2-7}%
& \multirow{2}{*}{Eq.~\eqref{eq:Rtm}} & $P=\pi$ & $0.00076 \pm 0.00485$ & $0.00070 \pm 0.00485$ & $0.00214 \pm 0.00486$ & $0.00067 \pm 0.00565$ \\ \cline{3-7}%
& & $P=K$ & $0.00992 \pm 0.01007$ & $0.00998 \pm 0.01008$ & $0.00848 \pm 0.01003$ & $0.01403 \pm 0.01244$ \\ \hline%
\multicolumn{3}{c}{\cellcolor{gray!20}\textbf{Scheme B}} & \multicolumn{4}{c}{ } \\ \cline{1-6}%
\multicolumn{3}{|c|}{$R_{\tau/e}$} & $0.00665 \pm 0.00381$ & $0.00634 \pm 0.00383$ & $0.01336 \pm 0.00480$ & \multicolumn{1}{c}{ }\\ \cline{1-6}%
\end{tabular}%
}}\\%
\subfloat[]{\label{tab:Vt4Sq}%
\fbox{%
\begin{tabular}{|c|c|c|c|c|c|c|}
\multicolumn{7}{c}{\textbf{Predicted values for $\modulus{V_{\tau 4}}^2$}}\\ \hline%
\multicolumn{3}{|c}{\cellcolor{gray!20}\textbf{Scheme A}} & \multicolumn{3}{c|}{ } & \cellcolor{gray!20}\textbf{Scheme C} \\ \cline{4-7}%
\multicolumn{3}{|c|}{ } & \multicolumn{3}{c|}{$R_{e/\mu}$} & $R_{\tau/e}$\\ \cline{4-7}%
\multicolumn{3}{|c|}{ } & \multirow{2}{*}{Eq.~\eqref{eq:R-e-m}} & \multicolumn{2}{c|}{Eq.~\eqref{eq:Rem}} & \multirow{2}{*}{Eq.~\eqref{eq:R-t-e}} \\ \cline{5-6}%
\multicolumn{3}{|c|}{ } & & $P=\pi$ & $P=K$ & \\ \hline%
\multirow{3}{*}{$R_{\tau/\mu}$} & \multicolumn{2}{c|}{Eq.~\eqref{eq:R-t-m}} & $0.00376 \pm 0.00294$ & $0.00370 \pm 0.00294$ & $0.00510 \pm 0.00300$ & $0.00376 \pm 0.00278$ \\ \cline{2-7}%
& \multirow{2}{*}{Eq.~\eqref{eq:Rtm}} & $P=\pi$ & $0.00670 \pm 0.00342$ & $0.00664 \pm 0.00342$ & $0.00806 \pm 0.00349$ & $0.00527 \pm 0.00292$ \\ \cline{3-7}%
& & $P=K$ & $0.01222 \pm 0.00582$ & $0.01216 \pm 0.00582$ & $0.01364 \pm 0.00589$ & $0.00821 \pm 0.00389$ \\ \hline%
\multicolumn{3}{c}{\cellcolor{gray!20}\textbf{Scheme B}} & \multicolumn{4}{c}{ } \\ \cline{1-6}%
\multicolumn{3}{|c|}{$R_{\tau/e}$} & $0.00376 \pm 0.00279$ & $0.00382 \pm 0.00279$ & $0.00244 \pm 0.00284$ & \multicolumn{1}{c}{ }\\ \cline{1-6}%
\end{tabular}%
}}%
\caption{Predicted values for $\modulus{V_{\ell 4}}^2$ (with $\ell=e,\mu,\tau$)
from weak decays following schemes A, B and C. The pairs of ratio observables
used in these predictions are $R_{\tau/\mu}$ and $R_{e/\mu}$ for scheme A,
$R_{e/\mu}$ and $R_{\tau/e}$ for scheme B and, finally, $R_{\tau/\mu}$ and
$R_{\tau/e}$ for scheme C. In all the schemes the observable $N_{\nu}$, the
number of light neutrinos, is used.}%
\label{tab:Vl4Sq}%
\end{table*}

\begin{table*}[hbtp]
\centering%
\subfloat[]{\label{tab:ampl-emu}
\fbox{%
\begin{tabular}{|c|c|c|c|c|c|c|}
\multicolumn{7}{c}{\textbf{Predicted values for $\sin^2 2\theta_{\mu e} = 4 \modulus{V_{\mu 4}}^2 \modulus{V_{e 4}}^2$}}\\ \hline%
\multicolumn{3}{|c}{\cellcolor{gray!20}\textbf{Scheme A}} & \multicolumn{3}{c|}{ } & \cellcolor{gray!20}\textbf{Scheme C} \\ \cline{4-7}%
\multicolumn{3}{|c|}{ } & \multicolumn{3}{c|}{$R_{e/\mu}$} & $R_{\tau/e}$\\ \cline{4-7}%
\multicolumn{3}{|c|}{ } & \multirow{2}{*}{Eq.~\eqref{eq:R-e-m}} & \multicolumn{2}{c|}{Eq.~\eqref{eq:Rem}} & \multirow{2}{*}{Eq.~\eqref{eq:R-t-e}} \\ \cline{5-6}%
\multicolumn{3}{|c|}{ } & & $P=\pi$ & $P=K$ & \\ \hline%
\multirow{3}{*}{$R_{\tau/\mu}$} & \multicolumn{2}{c|}{Eq.~\eqref{eq:R-t-m}} & $\left(2.76 \pm 1.80 \right) \times 10^{-4}$ & $\left(2.81 \pm 1.83\right) \times 10^{-4}$ & $\left(1.17 \pm 1.72\right) \times 10^{-4}$ & $\left(2.76 \pm 1.67\right) \times 10^{-4}$ \\ \cline{2-7}%
& \multirow{2}{*}{Eq.~\eqref{eq:Rtm}} & $P=\pi$ & $\left(0.14 \pm 0.89\right) \times 10^{-4}$ & $\left(0.14 \pm 0.95\right) \times 10^{-4}$ & $\left(0.19 \pm 0.68\right) \times 10^{-4}$ & $\left(0.32 \pm 2.68\right) \times 10^{-4}$ \\ \cline{3-7}%
& & $P=K$ & $\left(2.44 \pm 4.80\right) \times 10^{-4}$ & $\left(2.33 \pm 4.77\right) \times 10^{-4}$ & $\left(4.36 \pm 6.33\right) \times 10^{-4}$ & $\left(8.29 \pm 7.67\right) \times 10^{-4}$ \\ \hline%
\multicolumn{3}{c}{\cellcolor{gray!20}\textbf{Scheme B}} & \multicolumn{4}{c}{ } \\ \cline{1-6}%
\multicolumn{3}{|c|}{$R_{\tau/e}$} & $\left(2.76 \pm 1.74\right) \times 10^{-4}$ & $\left(2.64 \pm 1.75\right) \times 10^{-4}$ & $\left(4.84 \pm 2.30\right) \times 10^{-4}$ & \multicolumn{1}{c}{ }\\ \cline{1-6}%
\end{tabular}%
}}\\%
\subfloat[]{\label{tab:ampl-ee}
\fbox{%
\begin{tabular}{|c|c|c|c|c|c|c|}
\multicolumn{7}{c}{\textbf{Predicted values for $\sin^2 2\theta_{e e} = 4 \modulus{V_{e 4}}^2 \modulus{1- \modulus{V_{e 4}}^2}$}}\\ \hline%
\multicolumn{3}{|c}{\cellcolor{gray!20}\textbf{Scheme A}} & \multicolumn{3}{c|}{ } & \cellcolor{gray!20}\textbf{Scheme C} \\ \cline{4-7}%
\multicolumn{3}{|c|}{ } & \multicolumn{3}{c|}{$R_{e/\mu}$} & $R_{\tau/e}$\\ \cline{4-7}%
\multicolumn{3}{|c|}{ } & \multirow{2}{*}{Eq.~\eqref{eq:R-e-m}} & \multicolumn{2}{c|}{Eq.~\eqref{eq:Rem}} & \multirow{2}{*}{Eq.~\eqref{eq:R-t-e}} \\ \cline{5-6}%
\multicolumn{3}{|c|}{ } & & $P=\pi$ & $P=K$ & \\ \hline%
\multirow{3}{*}{$R_{\tau/\mu}$} & \multicolumn{2}{c|}{Eq.~\eqref{eq:R-t-m}} & $\left(4.10 \pm 1.67 \right) \times 10^{-2}$ & $\left(4.22 \pm 1.67\right) \times 10^{-2}$ & $\left(1.46 \pm 2.05\right) \times 10^{-2}$ & $\left( 4.10 \pm 1.08\right) \times 10^{-2}$ \\ \cline{2-7}%
& \multirow{2}{*}{Eq.~\eqref{eq:Rtm}} & $P=\pi$ & $\left(1.79 \pm 2.18\right) \times 10^{-2}$ & $\left(1.92 \pm 2.18\right) \times 10^{-2}$ & $\left(0.88 \pm 2.48\right) \times 10^{-2}$ & $\left(4.69 \pm 1.14\right) \times 10^{-2}$ \\ \cline{3-7}%
& & $P=K$ & $\left(2.44 \pm 4.10\right) \times 10^{-2}$ & $\left(2.32 \pm 4.10\right) \times 10^{-2}$ & $\left(5.08 \pm 4.20\right) \times 10^{-2}$ & $\left(5.82 \pm 1.50\right) \times 10^{-2}$ \\ \hline%
\multicolumn{3}{c}{\cellcolor{gray!20}\textbf{Scheme B}} & \multicolumn{4}{c}{ } \\ \cline{1-6}%
\multicolumn{3}{|c|}{$R_{\tau/e}$} & $\left(4.10 \pm 1.09\right) \times 10^{-2}$ & $\left(4.13 \pm 1.09\right) \times 10^{-2}$ & $\left(3.59 \pm 1.11\right) \times 10^{-2}$ & \multicolumn{1}{c}{ }\\ \cline{1-6}%
\end{tabular}%
}}\\%
\subfloat[]{\label{tab:ampl-mumu}
\fbox{%
\begin{tabular}{|c|c|c|c|c|c|c|}
\multicolumn{7}{c}{\textbf{Predicted values for $\sin^2 2\theta_{\mu \mu} = 4
		\modulus{V_{\mu 4}}^2 \modulus{1-\modulus{V_{\mu 4}}^2}$}}\\ \hline%
\multicolumn{3}{|c}{\cellcolor{gray!20}\textbf{Scheme A}} & \multicolumn{3}{c|}{ } & \cellcolor{gray!20}\textbf{Scheme C} \\ \cline{4-7}%
\multicolumn{3}{|c|}{ } & \multicolumn{3}{c|}{$R_{e/\mu}$} & $R_{\tau/e}$\\ \cline{4-7}%
\multicolumn{3}{|c|}{ } & \multirow{2}{*}{Eq.~\eqref{eq:R-e-m}} & \multicolumn{2}{c|}{Eq.~\eqref{eq:Rem}} & \multirow{2}{*}{Eq.~\eqref{eq:R-t-e}} \\ \cline{5-6}%
\multicolumn{3}{|c|}{ } & & $P=\pi$ & $P=K$ & \\ \hline%
\multirow{3}{*}{$R_{\tau/\mu}$} & \multicolumn{2}{c|}{Eq.~\eqref{eq:R-t-m}} & $\left(2.64 \pm 1.33\right) \times 10^{-2}$ & $\left(2.62 \pm 1.33\right) \times 10^{-2}$ & $\left(3.17 \pm 1.35\right) \times 10^{-2}$ & $\left(2.64 \pm 1.43\right) \times 10^{-2}$ \\ \cline{2-7}%
& \multirow{2}{*}{Eq.~\eqref{eq:Rtm}} & $P=\pi$ & $\left(0.31 \pm 1.94\right) \times 10^{-2}$ & $\left(0.28 \pm 1.94\right) \times 10^{-2}$ & $\left(0.85 \pm 1.94\right) \times 10^{-2}$ & $\left(0.27 \pm 2.26\right) \times 10^{-2}$ \\ \cline{3-7}%
& & $P=K$ & $\left(3.93 \pm 3.95\right) \times 10^{-2}$ & $\left(3.95 \pm 3.95\right) \times 10^{-2}$ & $\left(3.36 \pm 3.94\right) \times 10^{-2}$ & $\left(5.53 \pm 4.84\right) \times 10^{-2}$ \\ \hline%
\multicolumn{3}{c}{\cellcolor{gray!20}\textbf{Scheme B}} & \multicolumn{4}{c}{ } \\ \cline{1-6}%
\multicolumn{3}{|c|}{$R_{\tau/e}$} & $\left(2.64 \pm 1.50\right) \times 10^{-2}$ & $\left(2.52 \pm 1.51\right) \times 10^{-2}$ & $\left(5.27 \pm 1.87\right) \times 10^{-2}$ & \multicolumn{1}{c}{ }\\ \cline{1-6}%
\end{tabular}%
}}%
\caption{Values for $\sin^2 2\theta_{\mu e}$, $\sin^2 2\theta_{ee}$ and $\sin^2
2\theta_{\mu \mu}$ predicted by using the values of $\modulus{V_{e 4}}^2$ and
$\modulus{V_{\mu 4}}^2$ (see Tables~\ref{tab:Ve4Sq} and \ref{tab:Vm4Sq}
respectively) predicted from weak decays following schemes A, B and C.}%
\label{tab:amplitude}
\end{table*}

Using the values of ratio observables as shown in Eq.~\eqref{eq:ratio-values} we
can predict the values for $\modulus{V_{\ell 4}}^2$ (with $\ell=e,\mu,\tau$)
following schemes A, B and C. The predictions for $\modulus{V_{e4}}^2$,
$\modulus{V_{\mu 4}}^2$ and $\modulus{V_{\tau 4}}^2$ are tabulated in
Tables~\ref{tab:Ve4Sq}, \ref{tab:Vm4Sq} and \ref{tab:Vt4Sq} respectively. Taking
the average of these determined values we get,
\begin{subequations}
\begin{align}
\overline{\modulus{V_{e 4}}^2} &= \left(8.53 \pm 5.36\right) \times 10^{-3}, \\%
\overline{\modulus{V_{\mu 4}}^2} &= \left(6.73 \pm 5.94\right) \times
10^{-3},\\%
\overline{\modulus{V_{\tau 4}}^2} &= \left(6.62 \pm 3.65\right) \times 10^{-3}.
\end{align}
\end{subequations}
Once again these values are consistent with $0$, i.e.\ with no active-sterile
mixing, within $2\sigma$ standard deviation. The upper limits for
$\overline{\modulus{V_{e4}}^2}$, $\overline{\modulus{V_{\mu 4}}^2}$ and
$\overline{\modulus{V_{\tau 4}}^2}$ at $90\%$ confidence level are $1.73 \times
10^{-2}$, $1.65 \times 10^{-2}$ and $1.26 \times 10^{-2}$ respectively. From the
SBL global fits~\cite{SBL:global-analysis} we find that (i) $\modulus{V_{e4}}^2
\approx 0.010 \pm 0.003$ which is compatible with our estimate, (ii) depending
on the value of the squared mass difference $\Delta m^2_{41}$ the upper limit of
$\modulus{V_{\mu 4}}^2$ at $90\%$ C.L.\ varies in the range $[0.004,0.007]$
which is smaller than our estimate and (iii) $\modulus{V_{\tau 4}}^2 < 0.13$ (at
$90\%$ C.L.) which is larger than our estimate. Nevertheless, we also compute
the values for $\sin^2 2\theta_{\mu e}$, $\sin^2 2\theta_{ee}$ and $\sin^2
2\theta_{\mu\mu}$, which are listed in Table~\ref{tab:amplitude}. Taking the
average of the predicted values of Table~\ref{tab:amplitude} we get,
\begin{subequations}\label{eq:sinSq2Tht}
\begin{align}
\left.\sin^2 2\theta_{\mu e}\right|_{\text{avg}} &= \left(2.53 \pm
2.77\right)\times 10^{-4},\\%
\left.\sin^2 2\theta_{ee}\right|_{\text{avg}} &= \left(3.38 \pm
2.11\right)\times 10^{-2},\\%
\left.\sin^2 2\theta_{\mu\mu}\right|_{\text{avg}} &= \left(2.67 \pm
2.34\right)\times 10^{-2}.%
\end{align}
\end{subequations}
These results are consistent with $0$ within $2\sigma$ standard deviation. The
upper limits for $\sin^2 2\theta_{\mu e}$, $\sin^2 2\theta_{ee}$ and $\sin^2
2\theta_{\mu\mu}$ at $90\%$ confidence level are $7.07 \times 10^{-4}$, $6.84
\times 10^{-2}$ and $6.50 \times 10^{-2}$ respectively. Thus, no presence of
sterile neutrinos can be inferred from the existing data on charged-current weak
decays and the invisible decay width of the $Z$ boson. Nevertheless, it is very
interesting to note that the 2018 MiniBooNE result~\cite{MiniBooNE-anomaly}
hints at the possibility that there might exist a light sterile neutrino with
appreciable mixing with light active neutrinos with a best-fit value of $\sin^2
2\theta = 0.894$. This value is far above the value predicted in
Eq.~\eqref{eq:sinSq2Tht}. Thus our prediction directly contradicts the MiniBooNE
result. On the other hand, if MiniBooNE result were correct we should observe
the effect of the sterile neutrino in the weak decays we have discussed. We can
also compare our estimates with the best fit values given by SBL global fit
results\cite{SBL:global-analysis}
\begin{equation*}
\sin^2 2\theta_{\mu e} =
\begin{cases}
6.97 \times 10^{-3} & \text{(DaR)}\\
6.31 \times 10^{-3} & \text{(DiF)}
\end{cases},
\end{equation*}
where DaR and DiF denote the fact that the global fit includes neutrinos and
antineutrinos produced from $\pi$ decay-at-rest and $\pi$ decay-in-flight
respectively. These values are larger than our estimates.

\section{Conclusion}\label{sec:conclusion}

In conclusion we would like to emphasize that the approach we elaborated in this
paper can provide an independent and robust probe to active-sterile neutrino
mixing in addition to the traditional approach of using short-baseline neutrino
oscillation experiments. Using the precision measurements of the low energy
charged current processes, namely leptonic decays of $\mu$, $\tau$, $\pi$, $K$
and semi-leptonic decays of $\tau$ we define three ratio observables. Along with
these three ratio observables, which can be easily studied experimentally, we
also use the number of light neutrinos from the invisible decay of the $Z$ boson
which is also a very precise measurement. These four quantities form the basis
of our methodology. There are three numerical schemes for finding all the three
active-sterile mixing matrix elements, viz.\ $\modulus{V_{\ell 4}}$ for
$\ell=e,\mu,\tau$. If there exists a sterile neutrino having appreciable mixing
with active neutrinos, it would affect the precision measurements used in our
approach. Our approach, therefore, can be used not only to discover a sterile
neutrino, but also to study the mixing very precisely. It is also important to
note that our approach is strictly valid if the $4\times 4$ neutrino mixing
matrix is not unitary and if lepton universality is not imposed a priori. Both
the assumptions are not in conflict with currently existing experimental data.
If one considers the $4\times 4$ neutrino mixing matrix to be unitary, then the
non-oscillation observables considered in our approach become redundant and do
not constrain those mixings and the sterile neutrino hypothesis can not be
tested with our method. However, as is evident from our numerical results, the
$4\times4$ mixing matrix is consistent with being a unitary matrix. Finally we
must note that our numerical analysis considering the existing data is
consistent with no sterile neutrino hypothesis. Nevertheless, in the case of any
future claim of discovery of sterile neutrino from short-baseline neutrino
oscillation experiment, it would be necessary to test the discovery claim with
the method we have presented here.

\acknowledgments

This work of C.S.K.\ and D.S.\ was supported by the National Research Foundation
of Korea (NRF) grant funded by the Korean government (MSIP)
No.~2018R1A4A1025334. G.L.C.\ is grateful to Conacyt for financial support under
Project No.~236394. This work of D.S.\ was also supported (in part) by the
Yonsei University Research Fund (Post Doc.\ Researcher Supporting Program) of
2018 (project no.: 2018-12-0145). We would also to thank the anonymous referee
whose suggestions have improved our discussion of results and overall
presentation.

\end{document}